\documentclass[reprint,amsmath,amssymb,aps,pra,superscriptaddress,nofootinbib,twocolumn,floatfix]{revtex4-2}

\usepackage[USenglish]{babel}
\usepackage[T1]{fontenc}
\usepackage{hyperref}
\usepackage{graphicx}% Include figure files#

\usepackage{float}
\usepackage{placeins}
\usepackage{booktabs}
\usepackage{subcaption}
\usepackage{makecell}
\usepackage{tabstackengine}

\usepackage{dcolumn}% Align table columns on decimal point
\usepackage{bm}% bold math
\usepackage{braket}
\usepackage{xcolor}
\usepackage{bbold}
\usepackage{dsfont}
\usepackage{float}

\usepackage{xspace}
\usepackage{siunitx}
\usepackage{xfrac}
\usepackage{hyperref}
\usepackage[nameinlink]{cleveref}
\usepackage{appendix}

\usepackage{xifthen}
\usepackage{xcolor}
\usepackage{enumitem}
\hypersetup{
	colorlinks,
	linkcolor={red!75!black},
	citecolor={red!75!black},
	urlcolor={blue!75!black}
}

\usepackage[utf8]{inputenc}

\graphicspath{{figures/}}
\begin{document}

\preprint{APS/123-QED}

%make title
\title{Super-Resolving Normalising Flows for Lattice Field Theories}% Force line breaks with \\
\author{Marc Bauer}
\author{Renzo Kapust}
\email{kapust@thphys.uni-heidelberg.de}
\affiliation{
    Institute for Theoretical Physics, Universit\"{a}t Heidelberg,
Philosophenweg 16, D-69120, Germany
}

\author{Jan M. Pawlowski}
\affiliation{
    Institute for Theoretical Physics, Universit\"{a}t Heidelberg,
Philosophenweg 16, D-69120, Germany
}%
\affiliation{
    ExtreMe Matter Institute EMMI, GSI, Planckstr. 1, D-64291 Darmstadt, Germany
}

\author{Finn L. Temmen}
\affiliation{
    Institute for Theoretical Physics, Universit\"{a}t Heidelberg,
Philosophenweg 16, D-69120, Germany
}%

\date{\today}

\begin{abstract}
	We propose a renormalisation group inspired normalising flow that combines
benefits from traditional Markov chain Monte Carlo methods and standard
normalising flows to sample lattice field theories. Specifically, we use samples
from a coarse lattice field theory and learn a stochastic map to the targeted
fine theory. The devised architecture allows for systematic improvements and
efficient sampling on lattices as large as $128 \times 128$ in all phases when
only having sampling access on a $4\times 4$ lattice. This paves the way for
reaping the benefits of traditional MCMC methods on coarse lattices while using
normalising flows to learn transformations towards finer grids, aligning nicely
with the intuition of super-resolution tasks. Moreover, by optimising the base
distribution, this approach allows for further structural improvements besides
increasing the expressivity of the model.
\end{abstract}

\maketitle

%%%%%%%%%%%%%%%%%%%%%%%%%%%%%%%%%%%%%%%%%%%%
\section{Introduction}
\label{sec:Introduction}
Lattice field theory constitutes one of the primary methods for solving 
quantum field theories non-perturbatively. For sampling from the respective 
statistical integrals, traditional Markov chain Monte Carlo (MCMC) methods have
been effectively employed, providing valuable insights into the physics of
various many-body systems. 

The limitations of traditional MCMC methods, such as the need for individual
simulations for different points in the parameter space and critical slowing
down near a second-order phase transition, have prompted the exploration of new
approaches for sampling Boltzmann distributions.

Moreover, a crucial aspect in lattice field theory is the continuum limit which requires an 
access to finer and finer lattices for an extrapolation toward the
continuum theory of interest. As the continuum limit constitutes a second-order
phase transition itself and is inherently computationally expensive this
further motivates the search for more efficient algorithms.

With the advent of machine learning techniques in physics, normalising flows
have emerged as a prominent candidate to alleviate or transform the challenges
of traditional MC\-MC sampling approaches. While delivering impressive results,
normalising flows come with their own issues, including unfavourable volume
scaling of the complexity of the model and mode collapse in multimodal distributions.
 
In this work, we combine the benefits of traditional MCMC methods and
normalising flows rather than replacing one by the other. We do so by noting
that normalising flows have been used in the machine learning community to
tackle super-resolution tasks~\cite{Lugmayr:2020,Kothari:2021, Yao:2023}
effectively. There, one wants to obtain a high-resolution image from a
low-resolution one. In the context of lattice field theory, this translates to
learning an inverse renormalisation group step towards the continuum limit from
a given coarse lattice.
 
Recently, first steps to investigate multiscale normalising flows for a lattice
field theory as well as learning (inverse) renormalisation group
transformations have been made~\cite{Li:2018, 
Bachtis:2021eww, Hu:2020, Hou:2023, 
Abbott:2024knk, Albandea:2023, Mate:2024, Voleti:2024}. In this
paper, we propose a normalising flow architecture that connects a coarse and
fine lattice while incorporating intuitions of the renormalisation group. The
coarse lattice hereby stems from traditional MCMC algorithms while the
stochastic map to the finer lattice is learned by the flow. Importantly, we
also show how to compute optimal couplings in these situations.
 
In practice, we demonstrate the sampling of field theories on lattices as large
as $128 \times 128$ in the symmetric and broken phase when running an efficient MCMC method on a $4
\times 4$ system. Moreover, by starting on finer grids or further adapting the
base distribution, the methods presented here allow for a systematically
improvable architecture beyond making the learned transformations between the
grids more expressive.
 
The paper is structured as follows: In \Cref{sec:Phi4Theory}, we introduce the
lattice field theory setup and the $\phi^4$-theory as a canonical example. In
\Cref{sec:NormalizingFlows}, we introduce square and rectangular normalising
flows and discuss their application to lattice field theory. This enables us to
present the renormalisation group inspired normalising flows in
\Cref{sec:Architecture} and to discuss the used  optimisation criterion as well
as the  optimisation of the used couplings in \Cref{sec:OptimizationCriterion}.
We then showcase the newly modulated architecture in \Cref{sec:IRMatching},
where we fix the distribution on the fine lattice and then find the best
transformation and coarse lattice distribution for this sampling task. We
conclude in \Cref{sec:Conclusion}.

%%%%%%%%%%%%%%%%%%%%%%%%%%%%%%%%%%%%%%%%%%%%
\section{Scalar lattice field theory}
\label{sec:Phi4Theory}

In this work we only consider a scalar field theory, but we emphasise that our general approach is not restricted to it. For the scalar theory the Euclidean path
integral is given by 
\begin{align}
    \mathcal{Z} = \int \mathcal{D}\phi\, e^{-\mathcal{S}[\phi]}\,,
    \label{eq:PathIntegral}
\end{align}
where the action $\mathcal{S}[\phi]$ is a functional of the continuous field
$\phi(x)$. In the following we use the $\phi^4$-theory as an example to
illustrate the concepts and methods introduced in this work. Its action is
given by
\begin{align}
\mathcal{S}[\phi]\! = \!\int \! d^d\!x \left[ \frac{1}{2} \phi(x) \left(-\Delta +M^2 \right) \phi(x) + \frac{g}{4!} \phi^4(x) \right]\,,
\end{align}
with the mass $M$ and coupling $g$. In the lattice formulation, we discretise
the field $\phi(x)$ on a $d$ dimensional hypercubic lattice $\Lambda$ with $L$
lattice sites in each dimension, separated by the lattice distance $a$. We refer to $L$,
the number of grid points in each direction, as the \textit{linear lattice
size}, which should not be confused with the physical spacetime extent $l\, a$ of the
lattice. The discretised lattice action is then given by
\begin{align}
    \label{eq:ActionPhi4Theory}
    S(\phi) = \sum_{x\in \Lambda} \left[ ( 1 - 2 \lambda) \, \phi_x^2 + \lambda \phi_x^4 - 2\kappa \sum_\mu \phi_x \phi_{x+\mu} \right]\,.
\end{align}
Here, we have defined the dimensionless quantities $\kappa,\,\lambda$, and $\phi_x$ via
\begin{align}\nonumber 
	a^{\frac{d-2}{2}} \phi(x) =& \sqrt{2\kappa} \phi_x \\[1ex]\nonumber 
	\left(a M\right)^2 = &\frac{1-2\lambda}{\kappa} -2d\\[1ex] 
	a^{4-d} g = &\frac{6\lambda}{\kappa^2}\,,
	\label{eq:DimelessQuantities} 
\end{align}
where $x \in \Lambda$.
The continuum theory is retrieved in the limit of $a \to 0$, $L \to \infty$,
while keeping the simulated physics fixed.
In discretised form, we can think of the path integral as a high-dimensional
regular integral
\begin{align}
 Z = \int D\phi\,e^{-S(\phi)}\,, \quad D\phi = \left( \,\prod_{x \in \Lambda} d\phi_x \, \right)\,.
\end{align}
From the statistical physics point of view, the action induces a Boltzmann
distribution
\begin{align}
 p(\phi) = \frac{e^{-S(\phi)}}{Z} \,,
\end{align}
on the lattice, and observables $\mathcal{O}(\phi)$ are computed according to
\begin{align}
 \left\langle \mathcal{O} \right\rangle = \int D\phi \, \mathcal{O}(\phi) \,\frac{e^{-S(\phi)}}{Z}\,.
\end{align}
Interpreting this expression as a statistical integral enables us to compute
the expectation value using Monte Carlo algorithms. Markov chain based
approaches, where new samples only depend on its direct predecessor, are the
chief method of choice for most lattice computations to date. For a Markov
chain Monte Carlo algorithm to be efficient, it must have a low autocorrelation
time, which means that the correlation between configurations in the Markov
chain should decrease rapidly.

Unfortunately, close to second-order phase transitions, autocorrelation times diverge for any known
general MCMC algorithm, as the correlation
length diverges. This problem is called critical slowing
down~\cite{Wolff:1989wq}. Since the continuum limit constitutes a second-order
phase transition of the underlying statistical system, the critical slowing
down problem presents itself as a general issue in lattice simulations.

This concludes our quick introduction of the lattice field theory setup.

%%%%%%%%%%%%%%%%%%%%%%%%%%%%%%%%%%%%%%%%%%%%
\section{Normalising flows}
\label{sec:NormalizingFlows}

In this section, we discuss normalising flows, which are a general class of
machine learning models that allow for efficient sampling and probability
density estimation. In the lattice field theory context they were proposed as
an interesting alternative sampling method to tackle the critical slowing down
problem \cite{Albergo:2019, Kanwar:2020}.

Within normalising flows, one tries to find a transformation that maps
configurations from a distribution that is easy to sample to configurations
that are distributed according to the target distribution of the chosen lattice
field theory. Moreover, the flow should do so while efficiently estimating the
configurations' likelihood, allowing for exact Monte Carlo approaches.

In the following, we will go into more detail of this sampling approach and
introduce square and rectangular flows \cite{Caterini:2021} which will be
used to build the proposed architecture of this paper in
\Cref{sec:Architecture}.

%%%%%%%%%%%%%%%%%%%%%%%%%%%%%%%%%%%%%%%%%%%%
\subsection{Sampling with a normalising flow}\label{sec:NFIFBasics}

Assume we want to sample fields $\phi \in \mathds{R}^{\mathcal{L}^d}$ according
to the distribution $p_\mathcal{L}$ induced by the action $S_\mathcal{L}$ on
the lattice $\Lambda_\mathcal{L}$ with linear lattice size $\mathcal{L}$. We call this distribution the
target distribution.

To utilise a normalising flow for sampling from this distribution, we introduce
another field theory with the fields $\varphi \in \mathds{R}^{L^d}$ that we can
sample according to a distribution induced by the action $S_L$ on the lattice
$\Lambda_L$. This distribution is called prior or base
distribution, and we choose $\mathcal{L} \geq L$.

We now also introduce a map
\begin{align}
	\mathcal{T}_\theta: \mathds{R}^{L^d} &\to \mathds{R}^{\mathcal{L}^d}\,,
\end{align}
with learnable parameters $\theta$ that connects the fields on $\Lambda_L$ to
fields on $\Lambda_\mathcal{L}$. We call
\begin{align}
	\Tilde{\phi} = \mathcal{T}_\theta (\varphi)\,, \quad \log \Tilde{p}(\Tilde{\phi})\,,
\end{align}
the push forward of the map and its log-likelihood, respectively. The
normalising flow is expected to estimate both efficiently.

The intuition behind a sampling algorithm based on this setup is the
following: Let us assume that we have a base distribution from which we can easily
sample i.i.d configurations. Then, starting from these configurations, we
optimise the map $\mathcal{T}_\theta$ such that the push forward distribution
$\Tilde{p}$ approximates the target distribution $p_\mathcal{L}$ well. This
allows us to sample from the target distribution by first sampling from the
base distribution and then applying the transformation $\mathcal{T}_\theta$.
Accordingly, we have transformed the hard sampling problem of the target
distribution into the significantly easier one of the base distribution. The
costs we have to muster for this transfer are the training costs of the
learnable parameters $\theta$.

In many situations, one chooses a standard Gaussian with i.i.d. components as
the base distribution. Alternatively, a non--interacting field theory may also
be used. In this work, we will consider a plethora of potential base
distributions focusing on how they can be constructed and optimised from
coarser theories.
 
Needless to say, the optimisation of the transformation will generally not be
perfect. So, an accept-reject step with the acceptance probability
\begin{align}
    \label{eq:AcceptReject}
    \text{acc}(\phi_c, \Tilde{\phi}) = \min\left(1 , \frac{\Tilde{p}(\phi_c)\, p(\Tilde{\phi})}{p(\phi_c) \, \Tilde{p}(\Tilde{\phi})}\right)\,,
\end{align}
is necessary to ensure exact sampling in the target theory. Here, $\phi_c$
represents the current configuration in the Markov chain, and $\Tilde{\phi}$
represents the configuration proposed by the density-estimating model. Since
the total likelihood of each configuration is tracked as it progresses through
the flow, we can utilise the likelihood ratio in the acceptance probability.
 
To evaluate the quality of the normalising flow, we follow the literature in
considering two further quantities. Firstly, we look at the (reverse)
Kullback-Leibler divergence,
\begin{align}
 D_{KL}(\Tilde{p}\| p) = \int D\Tilde{\phi}\, \Tilde{p}(\Tilde{\phi}) \log \left( \frac{\Tilde{p}(\Tilde{\phi})}{p(\Tilde{\phi})} \right)\,,
\end{align}
which provides an (improper) measure of the difference between the push forward
distribution $\Tilde{p}$ and the target distribution $p$. This quantity is
useful because it can be estimated using only samples from the push forward
distribution. Secondly, we use the effective sample size (ESS), given by
\begin{align}
    \text{ESS}/N = \frac{1}{\int D\Tilde{\phi}\, \Tilde{p}(\Tilde{\phi}) \, w^2(\phi)} \approx \frac{\left( \sum_{i=1}^N w_i \right)^2}{\sum_{i=1}^N w_i^2}\,,
\end{align}
where we have defined $w_i = p(\Tilde{\phi}_i)/ \Tilde{p}(\Tilde{\phi}_i) $ for each of the $N$ samples~\cite{Hackett:2021idh}. 
While a good ESS/$N$ translates into a good acceptance rate \labelcref{eq:AcceptReject}, the ESS/$N$ has the
benefit that we must not set up a Markov chain to estimate it.
 
This concludes the general overview and structure of the sampling algorithm
built upon in this work.
 
Now, we discuss the implications of changing the dimensionality of the field when
pushing it through the transformation $\mathcal{T}_{\theta}$. For that matter,
we introduce square and rectangular normalising flows in the following section,
and discuss how to compute the (log)-likelihood in these situations.

%%%%%%%%%%%%%%%%%%%%%%%%%%%%%%%%%%%%%%%%%%%%
\subsection{Square normalising flows}
\label{sec:SquarNormalizingFlow}

For the square normalising flow, we consider the situation where
$\mathcal{L} = L$ such that we remain on an equally fine lattice. Moreover, we
take the map $\mathcal{T}_\theta$ to be bijective with the inverse
$\mathcal{T}_\theta^{-1}$. The transformation induces a change of variables
under the statistical integral starting from the base distribution
$p_L(\varphi)$, such that the density changes according to
\begin{align}
    \label{eq:ChangeOfVariables}
    \Tilde{p}(\Tilde{\phi}) = p_L(\varphi) \left| \det J_\mathcal{T}(\varphi)\right|^{-1} \quad \text{with} \quad \varphi = \mathcal{T}^{-1}_\theta(\Tilde{\phi})\,.
\end{align}
Here, $J_\mathcal{T} =  \partial\,\mathcal{T}_\theta(\varphi) /  \partial
\varphi$ denotes the Jacobian of the transformation $\mathcal{T}_\theta$. It is 
a square matrix since the transformation does not change the dimensionality
of the fields. The (log-)likelihood computation must be tractable, which is one
of the main challenges when constructing normalising flows. In particular,
efficient evaluations of the $\log \det J_\mathcal{T}$ term, where
$J_\mathcal{T}$ scales with the lattice volume, are imperative.
 
Square normalising flows are typically performant on relatively coarse lattices
but scale poorly when increasing the number of
sites~\cite{DelDebbio:2021,Abbott:2023,
Kumar:2020}. Moreover, when simulating lattice field theories in the
broken phase with its multi-modal distribution, some normalising flow
models struggle to perform well due to mode
collapse~\cite{Hackett:2021idh}. Lastly, in the vicinity of the phase
transition, the critical slowing down problem is generally not solved but
instead transferred into the learning process~\cite{DelDebbio:2021}.

%%%%%%%%%%%%%%%%%%%%%%%%%%%%%%%%%%%%%%%%%%%%
\subsection{Rectangular normalising flows}
\label{sec:RectangularNormalizingFlow}

In this paper, we aim to connect coarse and fine lattices with a normalising
flow. This alleviates some aforementioned issues, as the fine configurations
can inherit correlations from the used coarse samples. Moreover,
autocorrelation times are typically under control in coarse theories, and
efficient sampling is possible \cite{Bachtis:2021eww, Albandea:2023}.
 
Nevertheless, we cannot directly apply square normalising flows in these
situations. To connect the coarse lattice $\Lambda_L$ with linear lattice size
$L$ to the fine lattice $\Lambda_\mathcal{L}$ with linear lattice size
$\mathcal{L}$, where $\mathcal{L} > L$, we require a map
\begin{align}
    \mathcal{T}_\theta: \mathds{R}^{L^d} \to \mathds{R}^{\mathcal{L}^d}\,,
\end{align}
which maps the coarse field $\varphi$ to the fine field $\Tilde{\phi}$.
However, the change of variables formula \labelcref{eq:ChangeOfVariables} is
only applicable when the target and the base distribution relate to the same
number of lattice sites.
 
Instead, we consider the injective map $\mathcal{T}_\theta$ for which the more
general change of variables formula
\cite{Koethe:2023,Caterini:2021,Kumar:2020},
\begin{align}
    \label{eq:InjectiveChangeOfVariables}
    \Tilde{p}(\Tilde{\phi}) = p(\varphi) \left| \det J^T_\mathcal{T}(\varphi) \,J^{\phantom{T}}_\mathcal{T}(\varphi) \right|^{-1/2}\,,
\end{align}
with $ \varphi = \mathcal{T}^\dagger_\theta (\Tilde{\phi})$, is applicable.
Here, $\mathcal{T}_\theta^\dagger$ denotes the left inverse of the map
$\mathcal{T}_\theta$, such that
\begin{align}
    \label{eq:LeftInverse}
    \mathcal{T}_\theta^\dagger \circ \mathcal{T}^{\phantom{\dagger}}_\theta (\varphi) = \varphi\,.
\end{align}
Since we not only need the Jacobian determinant but also the matrix product
of the now non-square Jacobians, the likelihood computation is even more
demanding.
 
\begin{figure}[t]
	\centering
	\includegraphics[width=0.45\textwidth]{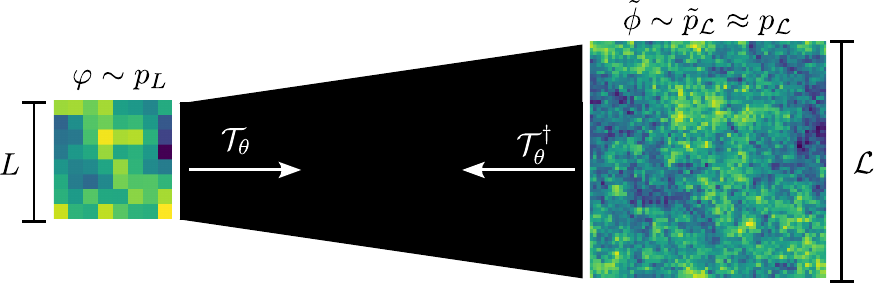}
	\caption{Visualization of the rectangular flow given by the optimisable transformations $\mathcal{T}_\theta: \mathds{R}^{L^d} \to \mathds{R}^{\mathcal{L}^d}$ and $\mathcal{T}^\dagger_\theta: \mathds{R}^{\mathcal{L}^d} \to \mathds{R}^{L^d}$. Here the coarse field $\varphi$ is distributed according to a Boltzmann distribution $p_L$ and the push forward $\Tilde{\phi}$ is thought to be approximately distributed according to the target Boltzmann distribution $p_\mathcal{L}$.\hspace*{\fill}  }
	\label{fig:Desideratum}
\end{figure}

This concludes the recap on square and rectangular normalising flows and leaves
us with the desired application depicted in \Cref{fig:Desideratum}. We show
the coarse lattice with configurations $\varphi$ distributed according to some
not necessarily trivial distribution $p_L$ and the map $\mathcal{T}_\theta$ to
the finer lattice configurations $\Tilde{\phi}$ which approximately follow the
target distribution $p_\mathcal{L}$.
 
Lastly, one can also see how the rectangular normalising flow relates to
learning block spinning transformations and their inverses. When
$\mathcal{T}_\theta^\dagger$ respects the symmetries of the lattice field
theory and maps to all possible configurations $\varphi$, it constitutes a
block spinning transformation. Then, the learned map $\mathcal{T}_\theta$ can,
in the appropriate sense, be thought of as an inverse renormalisation group
step.
 
In the following section, we will first show the proposed architecture
concretely and then discuss the invertibility of the transformation
$\mathcal{T}^\dagger_\theta$ further.
%

%%%%%%%%%%%%%%%%%%%%%%%%%%%%%%%%%%%%%%%%%%%%
\section{Architecture}
\label{sec:Architecture}

In this Section we discuss the details of constructing a normalising flow for the purpose
of sampling a lattice field theory on a fine lattice starting from a coarse
one. Our goal is to enable an efficient likelihood computation while
maintaining a solid connection to the renormalisation group.
 
As was already mentioned above, evaluating the change of variables \labelcref{eq:InjectiveChangeOfVariables} 
proves difficult in this situation as
we require $\det J_\mathcal{T}^T J^{\phantom{T}}_\mathcal{T}$ to be tractable.
Many approaches use Hutchinson samples to estimate this Jacobian
determinant~\cite{Sorrenson:2024,Kothari:2021,Caterini:2021}. However, in contradistinction to the
above approaches, the target distribution in a lattice field theory 
is already known, but we lack the tools to sample from it
efficiently. Additionally, we require the exact likelihood of each flowed
sample to construct a Markov chain on the target distribution and obtain exact
samples for the desired theory.
 
\begin{figure*}
	\centering
	\includegraphics[width=\textwidth]{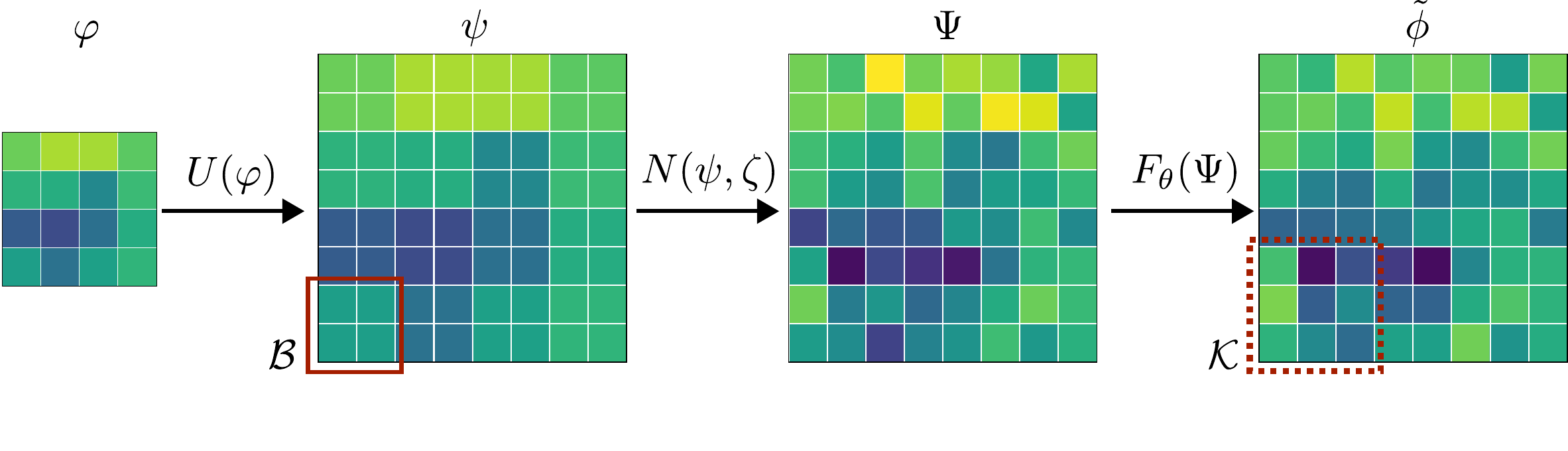}
	\caption{Illustration of the architecture of the flow $\mathcal{T}_\theta$. Given a coarse configuration $\varphi \in \mathds{R}^{L^d}$, the flow upsamples it na\"ively ($U$), applies Gaussian noise in an invertible manner ($N$), and pushes the noised configuration through a continuous normalising flow ($F_\theta$) with optimisable parameters $\theta$. Here $\mathcal{B}$ denotes the na\"ive upsampling blocks over which also the invertible noise is applied. Moreover, $\mathcal{K}$ denotes the local kernel of the continuous normalising flow.\hspace*{\fill}  }
	\label{fig:FlowExplained}
\end{figure*}

Here, we propose an architecture that instantiates the desired normalizing flow of
\Cref{fig:Desideratum}. It is close to the intuition of the renormalisation group 
and allows us to track the likelihood of the configurations throughout the flow
precisely. For this, we only assume that we have sampling access to the theory
on the coarse lattice and can compute the action on a fine lattice. The
introduced transformation maps the coarse configurations to a finer lattice by
which it enlarges the linear lattice size of the coarse lattice by a factor of
$b=2$ as shown in \Cref{fig:FlowExplained}.
 
Each building block will be explained in the following paragraphs. Moreover, we
will show how the log-likelihood can be computed precisely at each step and
comment on the notion of their invertibility.

%%%%%%%%%%%%%%%%%%%%%%%%%%%%%%%%%%%%%%%%%%%%
\subsection{Na\"ive upsampling}
\label{sec:ArchitectureNaiveUpsampling}

To retain the correlations embedded in the coarse configurations and to ensure
a tractable Jacobian determinant of the rectangular normalising flow, we
na\"ively upsample the coarse configurations, following the example
in~\cite{SR-Ising}. From the fields $\varphi_x$ on each coarse lattice site, we create a
$b^d$ block $\mathcal{B}$ of sites $\psi_{x\in \mathcal{B}} = \varphi_x$ on the
finer level. Having reshaped the data into a $L^d$ dimensional vector, the
Jacobian of this transformation is given by
\begin{align}
	J_{U} = \begin{pmatrix} 1 & 0 & \cdots & 0 \\ 
							\vdots & 0 & \cdots & 0 \\ 
							1 & 0 & \cdots & 0 \\ 
							0 & 1 & \cdots & 0 \\
							0 & \vdots & \cdots & 0 \\
							0 & 1 & \cdots & 0 \\
							\vdots &  &  & \vdots \\
							0 & 0 & \cdots & 1 \\
							0 & 0 & \cdots & 1
            \end{pmatrix}_{(bL)^d\times L^d}\,.
\end{align}
The Jacobian determinant reduces to a constant which is straightforwardly
computed as  $\log |\det J_U^T J^{\phantom{T}}_U | = L^d \log(b^d)$. By this
construction, we choose the left inverse $U^\dagger$ of the transformation $U$
to be given as the average over the blocks $\mathcal{B}$ used in the na\"ive
upsampling. This relates the operation to a standard block spinning
transformation. Accordingly, the new fields and their log-likelihood after the
na\"ive upsampling are given by
\begin{align}\nonumber 
		\psi = U(\varphi) \quad \, &, \quad \varphi = U^\dagger(\psi)\\[1ex]
	\log p_U(\psi) = \log p_L(&\varphi) - \frac{1}{2} L^d \log(b^d)\,.
	 \label{eq:NaiveUpsampling}
\end{align}

%%%%%%%%%%%%%%%%%%%%%%%%%%%%%%%%%%%%%%%%%%%%
\subsection{Adding invertible noise}
\label{sec:ArchitectureInvertibleNoise}

In general many configurations on the fine lattice will lead to the same
configuration on the coarse lattice. Moreover, starting from the coarse
lattice, we have no information on the fine (UV) degrees of freedom. This
motivates the addition of noise, $\zeta$. It ensures that the configuration
$\Tilde{\phi}$, we map to, is not solely determined by the coarse (IR) degrees of
freedom, $\varphi$, but also acknowledges the missing information from the UV.
Furthermore, inflating the configuration space with noise already finds many
applications in the context of normalising flows~\cite{Mate:2023,
Hovrat:2021, Hu:2020}.
 
To allow for the invertibility of the transformation in the way explained in
\Cref{sec:ArchitectureInvertibility}, we add the noise to each upsampling block
$\mathcal{B}$ such that it sums up to zero. This is easily achieved by sampling
the noise variables $\zeta_{\mathcal{B}, i}$ (with $i=1,\ldots, b^d-1$) of each
block from a multivariate Gaussian distribution with covariance matrix
\begin{align}
    \Sigma = \sigma^2_\theta\left(\mathds{I}_{b^d-1} - \frac{1}{b^d} \right)\,,
\end{align}
where $\mathds{I}_{b^d-1}$ is the $b^d-1$ dimensional identity matrix and
$\sigma_\theta$ is a learnable parameter. The final noise component is
determined by the averaging constraint, such that
\begin{align}\label{eq:NoiseConstraint}
    \zeta_{\mathcal{B}, b^d} = -\sum_{i=1}^{b^d-1} \zeta_{\mathcal{B}, i}\,.
\end{align}
Since the noise is sampled independently of $\psi$, the new fields and their
log-likelihood are given by
\begin{align}\nonumber 
    \Psi = &\,\psi + \zeta\,, \qquad \qquad \varphi =\,U^\dagger(\Psi)\,,\\[1ex]
    \log p_N(\Psi) = &\,\log p_U(\psi) + \log p_\zeta(\zeta)\,.
\label{eq:InvertibleNoiseAddition}
\end{align}
Here, $p_\zeta$ denotes the distribution of the noise. Furthermore, 
$U^\dagger$ also constitutes the left inverse of the flow after the noise
addition, $(N\circ U)^\dagger = U^\dagger$. This originates in the fact that the 
noise vanishes when taking the mean over the upsampling blocks $\mathcal{B}$,
 
Note that the coarse degrees of freedom $\varphi$ stem from $\mathds{R}^{L^d}$
and the noise degrees of freedom for  $\zeta$ come from
$\mathds{R}^{\mathcal{L}^d - L^d}$, as we ensure that the mean of the noise
over the blocks is zero. Accordingly, the embedding of the coarse lattice
configuration on the finer lattice and the noise ensures that the configuration
$\Psi$ spans the full $\mathds{R}^{\mathcal{L}^d}$ space. In particular, they do not only span 
the subspace of configurations that are compatible with the coarse lattice after
the upsampling.

%%%%%%%%%%%%%%%%%%%%%%%%%%%%%%%%%%%%%%%%%%%%
\subsection{Normalising flow}
\label{sec:ArchitectureNormalizingFlow}

We optimise a square normalising flow on the finer lattice
$\Lambda_\mathcal{L}$ for mapping this upsampled and noised configuration $\Psi \in
\mathds{R}^{\mathcal{L}^d}$ to a configuration approximately following a target
distribution. A similar approach to first embed the coarse degrees of
freedoms into a larger space and then apply a normalising flow can also be
found in~\cite{Brendan:2021,Hovrat:2021}.
 
For their recent success we choose continuous normalising flows (CNF) for this
task. A CNF is defined by a neural ordinary differential equation
(NODE)~\cite{Chen:2018} of the form
\begin{align}\label{eq:CNF_ODE}
	\frac{d \Psi(t)_x}{d t} = G^{(\mathcal{K})}_\theta(\Psi(t), t)_x \,.
\end{align}
Here $G^{(\mathcal{K})}_\theta(\Psi(t), t)_x$ is a function with the kernel
$\mathcal{K}$ of the field $\Psi(t)$ during the fictitious time $t$ with
optimisable parameters $\theta$. The great benefit of CNFs relies on the fact
that one can implement symmetries of the theories directly into the flow and
has great flexibility in the choice of $G^{(\mathcal{K})}_\theta$. We
align our construction of $G^{(\mathcal{K})}_\theta$ closely to the proposed
architecture in~\cite{Gerdes:2023} and parametrise $G^{(\mathcal{K})}_\theta$ as
\begin{align}
	G^{(\mathcal{K})}_\theta(\Psi(t), t)_x = \sum_{y, d, f} W_{x y d f} K(t)_d H(\, \Psi_y(t)\,)_f\,.
\end{align}
Here, $W$ denotes a learnable weight matrix, $K(t)$ a time kernel, and $H$ a
basis expansion of the field $\Psi_y(t)$. To ensure the $\mathds{Z}_2$ symmetry
of the $\phi^4$-theory, we choose the expansion
\begin{align}
H_1(x) = x\,, \qquad H_f(x) = \sin (\,\omega_f x\,)\,,
\end{align}
with learnable frequencies $\omega_f$, $f = 2 \ldots F$. For the time kernel
$K(t)_d$ we choose the first $D$ terms of the Fourier expansion on the time
interval $[t_1, t_2]$ we integrate over when solving the differential equation
\labelcref{eq:CNF_ODE}. $F$ and $D$ are hyperparameters of the CNF.
 
The lattice index $y$ only runs over the sites inside the kernel $\mathcal{K}$
(see \Cref{fig:FlowExplained}). This entails that the flow can only
directly couple sites in the scope of the kernel instantaneously, which reduces the number of
learnable parameters. We choose to do so since the coarse correlations should
already be present in $\Psi$. Hence, the flow must only learn the
short-scale correlations on the finer lattice.
 
As noted in~\cite{Gerdes:2023}, better training is achieved when we parametrise
the learnable weight matrix $W$ as
\begin{align}
	W_{x y d f} = \sum_{d' f'} \Tilde{W}_{x y d' f'} W^K_{d' d} W^H_{f' f}\,,
\end{align}
where the indices $d', f'$ run from one up to the bond dimensions $F', D'$ in
time and frequency space, respectively. $F'$ and $D'$ also are hyperparameters
of the CNF.
 
Furthermore, we identify spatial bonds ($x,y$) in $W$ that are equal under
the lattice symmetry. This further reduces the number of learnable parameters
and force the flow to respect symmetries present in the lattice field theory.
 
As we consider a $\phi^4$-theory on a periodic two-dimensional lattice, we ensure
that we respect the symmetry under $90^\circ$ rotations and mirror reflections.
Moreover, because of the upsampling in $b^d$ blocks, the field $\Psi$ admits to
a $b$-step translational symmetry.
 
At the beginning of the training the flow is initialised at the identity, which
means that at first no short scale correlations are added to the field, and
they enter over the training. In each training step, the NODE can then be
solved via the adjoint sensitivity method~\cite{Chen:2018,Gerdes:2023} with the
initial conditions $\Psi(t_0) = \Psi$, such that
\begin{align}
    \label{eq:FlowApplication}
	\Tilde{\phi} = F_\theta(\Psi) =  \Psi + \int_{t_0}^{t_1} d t\, G^{(\mathcal{K})}_\theta(\Psi(t), t)\,.
\end{align}
This is clearly invertible by changing the time direction of
integration. Moreover, the log--likelihood of the final configurations can be
computed by~\cite{Chen:2018}
\begin{align}
	\frac{d \log p_{F}(\Psi)}{d t} = - \nabla_\Psi \cdot G^{(\mathcal{K})}_\theta(\Psi(t), t)\,.
\end{align}
Accordingly, after the application of the flow the final log-likelihood is
given by
\begin{align}
	\log p_F(\Tilde{\phi}) = \log p_N(\Psi) - \int_{t_0}^{t_1} d t\, \nabla_\Psi \cdot G^{(\mathcal{K})}_\theta(\Psi(t), t)\,.
\end{align}
In analogy to discrete normalising flows, we will denote all the contributions
to the log-likelihood by the transformation $\mathcal{T}_\theta$ by `$\log \det
J_\mathcal{T}(\varphi)$'. Hence, the log-likelihood of the full transformation
$\mathcal{T}_\theta$ is given by
\begin{align}
	\log \Tilde{p}(\Tilde{\phi}) = \log p_L(\varphi) - \log \det J_\mathcal{T}(\varphi)\,, \quad \varphi = \mathcal{T}^\dagger_\theta(\Tilde{\phi})\,,
\end{align}
with $\mathcal{T}^\dagger_\theta (\Tilde{\phi}) = U^\dagger \circ
F^{-1}_\theta(\Tilde{\phi})$.
This concludes the direct discussion of each element of the architecture and
their interplay. To summarise, the proposed architecture na\"ively upsamples
the coarse configurations, adds noise, and then flows on the finer lattice. For
each step, the log-likelihood can be computed exactly and efficiently.
Furthermore, $\mathcal{T}_\theta: (\varphi; \zeta) \mapsto \Tilde{\phi}$ and
$\mathcal{T}_\theta^\dagger: \Tilde{\phi} \mapsto \varphi$ are both fixed
simultaneously. As a last step, we now also discuss the notion of invertibility
for the transformation $\mathcal{T}_\theta$.

%%%%%%%%%%%%%%%%%%%%%%%%%%%%%%%%%%%%%%%%%%%%
\subsection{Invertibility}
\label{sec:ArchitectureInvertibility}

By investigating the proposed architecture, we note that while
$\mathcal{T}_\theta^\dagger$ maps to a lower dimensional space, given
\textit{any} configuration $\phi \in \mathds{R}^{\mathcal{L}^d}$, the coarse as
well as noise degrees of freedom are always implicitly determined via
\begin{align}\nonumber 
    \varphi(\phi) &= \mathcal{T}^\dagger_\theta(\phi) \,\\[1ex]
    \zeta(\phi) &= \Psi(\phi) - \psi(\phi) = F^{-1}_\theta(\phi) - U(\,\mathcal{T}^\dagger_\theta(\phi)\,)\,.
\end{align}
This is in analogy to the work
in~\cite{Hu:2020,Hovrat:2021}, where the coarse and noise
degrees of freedom are explicitly separated.
 
It also nicely fits to the intuition that renormalisation group transformations
are not invertible because we loose information about the UV degrees of
freedom. Accordingly, this architecture manifests that the transformation would indeed be invertible (as
similarly touched upon in~\cite{Hu:2020, Cotler:2023}), if we would keep the UV
degrees of freedom exactly. Here, in
practice, we do not keep the noise degrees of freedom exactly but fix the
distribution they stem from.
 
Accordingly, although we have defined a left inverse
$\mathcal{T}_\theta^\dagger$ that parametrises a block spinning transformation,
it does so in the context of a bijective map on the finer lattice. So, this
architecture parametrises manifestly invertible transformations between
$(\varphi, \zeta) \in \mathds{R}^{L^d} \times \mathds{R}^{(\mathcal{L}^d -
L^d)}$ and $\phi \in \mathds{R}^{\mathcal{L}^d}$. Given a configuration
$\varphi$ on the coarse lattice, we know the subspace of $R^{\mathcal{L}^d}$ it
can map to when embedded with the additional noise $\zeta$. Moreover, given any
configuration $\phi$ on the fine lattice, we can uniquely determine a coarse
and a noise configuration $\varphi, \zeta$, from which it fictitiously stemmed.
This way, we paid tribute to the fact that the renormalisation group is a
semigroup while maintaining the invertibility of the transformation. More details 
on the interpretation of $\mathcal{T}^\dagger_\theta$ as a block spinning
transformation can be found in \Cref{sec:BlockSpinningTransformation}. 
 
This concludes the description of the proposed architecture. When discussing the  optimisation criterion in the
next section, we also consider the freedom of choosing the base and target
action and related applications.

%%%%%%%%%%%%%%%%%%%%%%%%%%%%%%%%%%%%%%%%%%%%
%%%%%%%%%%%%%%%%%%%%%%%%%%%%%%%%%%%%%%%%%%%%
\section{Optimization criterion}
\label{sec:OptimizationCriterion}

We have arrived at a normalising flow architecture that can relate coarse and
fine lattices. Our focus is now on the distributions induced by the actions
$S_L$ and $S_\mathcal{L}$ on the respective lattices, which shall represent
different discretisations of (roughly) the same continuum theory. With this, we follow an intuition also mentioned in~\cite{Albandea:2023ais}.
 
We aim at optimising the described transformation $\mathcal{T}_\theta$ and the
action on the coarse or fine lattice such, that the acceptance rate of the
Markov chain on the fine grid is maximised.
 
Concerning the couplings, this section does not further specify whether we want
to optimise the couplings of the base or target distribution. Whereas in
\Cref{sec:IRMatching}, we focus on the optimal couplings on the coarse lattice
for a given sampling problem on a target lattice, in \Cref{sec:UVMatching}, we
also discuss how one can sensibly adapt the couplings of the target distribution. 
 
To this end, we consider the reverse Kullback-Leibler divergence between the
target and the push forward distribution, $p_\mathcal{L}$ and
$\Tilde{p}_\mathcal{L}$ respectively, 
\begin{align}\nonumber 
 D_{KL}(\Tilde{p}_\mathcal{L} \| p_\mathcal{L}) = &\,\int D\Tilde{\phi}\, \Tilde{p}_\mathcal{L}(\Tilde{\phi}) \log \frac{\Tilde{p}_\mathcal{L}(\Tilde{\phi})}{p_\mathcal{L}(\Tilde{\phi})} \\[1ex]
    = &\,D_{KL}^{(I)}(\Tilde{p}_\mathcal{L} \| p_\mathcal{L}) + D_{KL}^{(II)}(\Tilde{p}_\mathcal{L} \| p_\mathcal{L})\,.
\label{eq:ReverseKLDivergence}
\end{align}
We have split the expression in \labelcref{eq:ReverseKLDivergence} into two terms: the first term holds the
part of the divergence that relates to an expectation value over the push
forward distribution,
\begin{align}\nonumber 
 D_{KL}^{(I)}(\Tilde{p}_\mathcal{L} \| p_\mathcal{L}) := \underset{\Tilde{\phi} \sim \Tilde{p}_\mathcal{L}}{\mathds{E}} [ S_\mathcal{L}(\Tilde{\phi}) &- S_L(\,\varphi(\Tilde{\phi})\,) \\[1ex]
    &- \log \det J_\mathcal{T}(\,\varphi(\Tilde{\phi})\,) ]\,. \label{eq:DKLI}
\end{align}
The second term holds the part of the divergence that relates to the
partition sums
\begin{align}
 D_{KL}^{(II)}(\Tilde{p}_\mathcal{L} \| p_\mathcal{L}) := \log Z_\mathcal{L} - \log Z_L\,.
\end{align}
For most applications, one effectively drops the $D_{KL}^{(II)}$ term in
\labelcref{eq:ReverseKLDivergence}: it is constant and irrelevant to the
 optimisation as long as the normalizations for the base and target
distributions remain fixed. However, we want to tune the base or target
distribution of the flow, making an  optimisation of the $D_{KL}^{(II)}$ term
necessary.
 
After all, in lattice field theory, we think of the lattice action
$S[c](\varphi)$ with bare couplings $c$ and the field $\varphi$ as a
discretised version of the continuum action $\mathcal{S}[\mathcal{C}][\phi]$
with the couplings $\mathcal{C}$ and the continuum field $\phi(x)$. The
continuum action is approached by the lattice action in the continuum limit and
the lattice action is structurally given in the same way for any lattice
$\Lambda$ and couplings $c$, see \labelcref{eq:ActionPhi4Theory}.
 
This leads us to the following consideration. Our flow connects two fields on different 
lattices $\Lambda_L$ and $\Lambda_\mathcal{L}$. Therefore, it is only reasonable to wish
for the distributions on the two lattices to be induced by different
discretisations $S_L[c_L](\varphi)$ and $S_\mathcal{L}[c_\mathcal{L}](\phi)$
given by structurally the same lattice action formula, for instance
\labelcref{eq:ActionPhi4Theory}.
 
We know from the renormalisation group, that the lattice couplings change under
this flow in a non-trivial way. Therefore, we either introduce the couplings of the base or
target actions as learnable parameters.
 
While the full partition sum is intractable for training, practically, we are
only interested in its gradient w.r.t. the training parameters (i.e. the
couplings).
 
Expressing the dependence of the partition sum on the learnable couplings $c_\theta$
as $Z[c_\theta]$, we can calculate the necessary gradient. Here we use insights from
the work on restricted Boltzmann machines and contrastive divergence
\cite{Hinton:2005}, 
\begin{align}\nonumber 
       \nabla_\theta \log Z[c_\theta] =&\, \frac{1}{Z[c_\theta]} \nabla_\theta Z[c_\theta] \\[1ex]\nonumber 
        =&\, \frac{1}{Z[c_\theta]} \nabla_\theta \int D\phi\, e^{-S[c_\theta](\phi)}  \\[1ex]
        = &\,- \underset{\phi \sim p[c_\theta]}{\mathds{E}}\left[\, \nabla_\theta S[c_\theta\, ](\phi)\,\right].
   \label{eq:logZGradient}
 \end{align}
The remaining gradient with respect to the learnable parameters is readily
computed. For most actions of interest, the number of couplings is small. Then, 
the expectation values can be reliably estimated using Monte Carlo techniques
that require only a modest number of samples, rendering the direct
computation of the gradient affordable during training.
 
In the $\phi^4$-theory investigated here, the gradients are given by
\begin{align}\nonumber 
        \frac{ \partial S}{ \partial \kappa_\theta} =&\, -2 \sum_{\substack{x\in \Lambda \\ \mu}} \phi_x \phi_{x+\mu} \,,\\[1ex]
        \frac{ \partial S}{ \partial \lambda_\theta} = &\,\sum_{x\in \Lambda} \left( \phi_x^4 -2\phi^2_x\right)\,.
\end{align}
With this, we are not only able to tune the transformation between the coarse
and fine action but also the couplings of the actions
themselves. Furthermore, one can also consider other, more general,
actions on the coarse lattice that allows for terms not present in the target
fine action. This is left to future work. 
 
During training we will use two sets of Monte Carlo samples: one set which
we push through the network to compute the gradient of $D_{KL}^{(I)}$. The other 
set is never pushed through the network but is used to compute the
gradient of $D_{KL}^{(II)}$ via \labelcref{eq:logZGradient}.
 
Here, we can already note that the couplings, for which we compute the
$D_{KL}^{(II)}$ gradient, do not change too much between each training step.
Moreover, since the respective set of samples is never pushed through the
normalising flow, we can safely reuse them.  Accordingly, one can use
reweighing techniques to significantly reduce the number of required Monte
Carlo simulations during training. For a given set of Monte Carlo samples with
couplings $c_\theta$, the $\log Z$ gradient with couplings $c'_\theta$ at a
later time in training can be estimated by
\begin{align}
    \nabla_\theta \log Z[c'_\theta] = -\frac{\underset{\phi \sim p[c_\theta]}{\mathds{E}}\left[\, \left(\,\nabla_\theta S[c'_\theta]\, \right) \frac{e^{-S[c'_\theta](\phi)}}{e^{-S[c_\theta](\phi)}}\, \right]}{\underset{\phi \sim p[c_\theta]}{\mathds{E}}\left[\, \frac{e^{-S[c'_\theta](\phi)}}{e^{-S[c_\theta](\phi)}}\, \right]}\,,
 \label{eq:Reweighing}
 \end{align}
The above procedure is applicable if the distributions $p[c_\theta]$ and $p[c'_\theta]$ are sufficiently close to each
other. Since the couplings do not change too much during training and
eventually converge at the end of it, the overlap between the distributions
$p[c_\theta]$ and $p[c'_\theta]$ at the two training times is typically quite
large.
 
This concludes our discussion of the  optimisation criterion.

%%%%%%%%%%%%%%%%%%%%%%%%%%%%%%%%%%%%%%%%%%%%
\section{IR-Matching}
\label{sec:IRMatching}

\begin{figure}[t]
	\centering
	\includegraphics[width=0.45\textwidth]{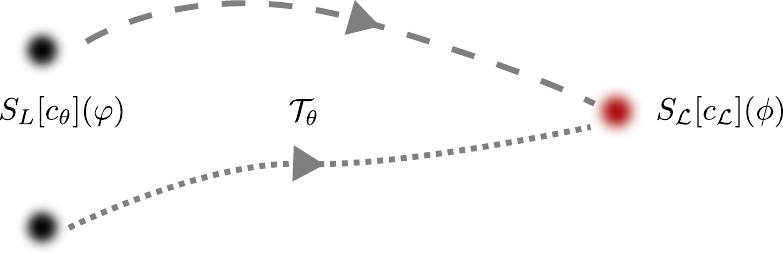}
	\caption{Illustration of the IR-Matching. The field $\varphi$ on the coarse lattice with action $S_L[c_\theta]$ and optimisable couplings $c_\theta$ is connected to the field $\phi$ on the fine lattice with action $S_\mathcal{L}[c_\mathcal{L}]$ and fixed couplings $c_\mathcal{L}$ by the transformation $\mathcal{T}_\theta$. \hspace*{\fill}  }
	\label{fig:IRMatching}
\end{figure}

In this Section we discuss an application of the proposed architecture and
 optimisation criterion, which we coin \textit{IR-Matching} and is illustrated
in \Cref{fig:IRMatching}. Concretely, we fix the  action
$S_\mathcal{L}[c_\mathcal{L}]$ on the fine lattice with couplings
$c_\mathcal{L}$ and find the optimal transformation $\mathcal{T_\theta}$ and
coarse couplings $c_\theta$ needed to sample from the target distribution.
 
First we will discuss more details of the approach. Then its
applicability in different phases for differently fine target lattices is considered. Details
on the used hyperparameters and error estimation can be found in
\Cref{sec:HyperparametersAndErrors}.

%%%%%%%%%%%%%%%%%%%%%%%%%%
\subsection{Details of IR-Matching }
\label{sec:IRMatchingMethodExplanation}

At the beginning of training, the couplings of both lattices have the same
value. Then, during training, the chosen couplings on the coarse lattice move
to the values optimal for the sampling problem at hand.
 
As the coarse couplings enter the sampling process of the coarse configurations
$\varphi$, we must use a sampling algorithm on the coarse lattice that is
differentiable w.r.t the learnable couplings. Here, we use a standard batched
Langevin sampling algorithm, where the fields follow the Langevin dynamics in
the fictitious time $\vartheta$ according to
\begin{align}\label{eq:Langevin}
    \varphi^{(\vartheta+1)} = \varphi^{(\vartheta)} -\tau \nabla_{\varphi} S[c_\theta](\varphi^{(\vartheta)}) + \sqrt{2\tau}\,\eta(\vartheta)\,,
\end{align}
where $\eta(\vartheta)$ denotes a Gaussian noise term and $\tau$ the Langevin
time step. We thermalise the set of coarse lattice configurations once and only
re-thermalise for the next training step with a slightly different set of
couplings. We note in this context that there are more sophisticated methods that also
utilise the adjoint method for the Langevin stochastic differential
equation~\cite{Li:2020, Alvestad:2023}. However, we found it sufficient for our purposes to simply 
backpropagate through the Langevin process. One of the reasons is that the lattices we sample from are quite small, and hence the memory cost required for the
backpropagation was not too demanding.

To sample on the fine lattice, we draw coarse configurations $\varphi$ as
described above and apply transformations $\mathcal{T}_{\theta_i}$ with $i = 1,
\ldots, \log_b(\mathcal{L}/L)$ with different weights $\theta_i$ successively. This 
yields
\begin{align}
    \Tilde{\phi} = \mathcal{T}_\theta(\varphi) = \mathcal{T}_{\theta_{\log_b(\mathcal{L}/L)}} \circ \ldots \circ \mathcal{T}_{\theta_1}(\varphi)\,,
\end{align}
belonging to the fine lattice $\Lambda_\mathcal{L}$. Each application of
$\mathcal{T}_{\theta_i}$ thereby doubles the linear lattice size of the lattice. 
The samples $\Tilde{\phi}$ on the fine lattice are then used to compute the $D_{KL}^{(I)}$ gradient of the loss from \labelcref{eq:DKLI}. Notably, the $D_{KL}^{(II)}$ gradient from
\labelcref{eq:logZGradient} only requires Monte Carlo samples on the coarse
lattice $\Lambda_L$, rendering this gradient estimation computationally
efficient.

\begin{figure*}[t]
    \centering
	\begin{subfigure}[t]{0.45\textwidth}
		\centering
		\includegraphics[width=\textwidth]{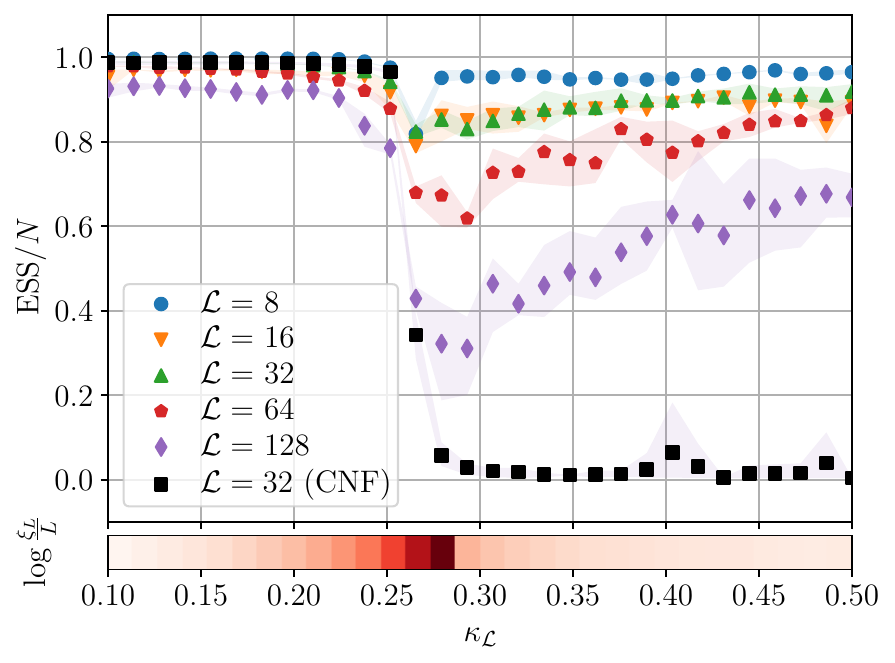}
		\caption{ESS/$N$ for the IR-Matching method on differently fine lattices and a CNF with $\mathcal{L}=32$ for comparison. \hspace*{\fill}  }
		\label{fig:IRMatchingDifferentPhasesESS}
	\end{subfigure}
	\hspace{0.05\textwidth}
	\begin{subfigure}[t]{0.45\textwidth}
		\includegraphics[width=\textwidth]{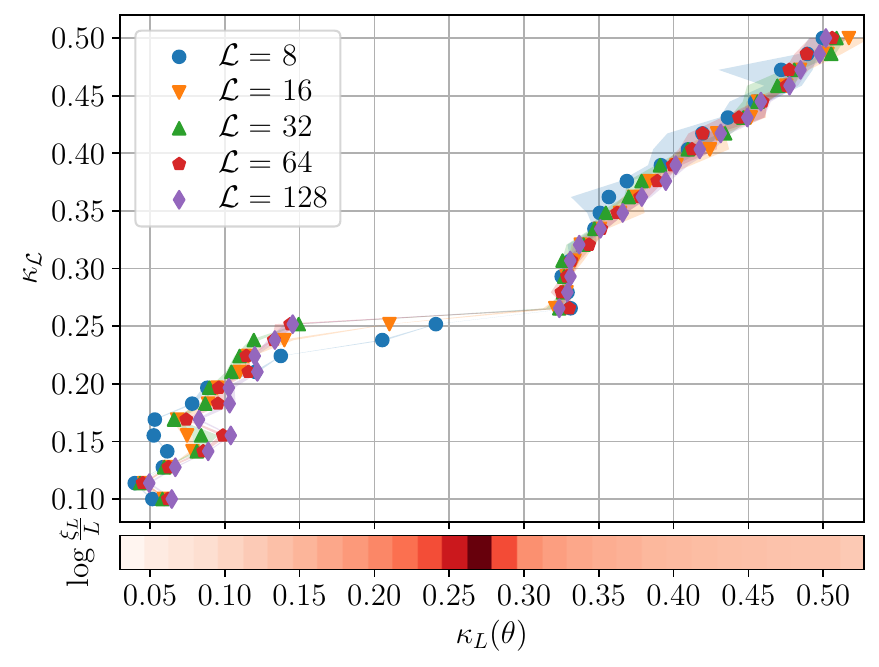}
		\caption{Optimizable coupling $\kappa_L(\theta)$ on the coarse lattice relative to the fixed couplings $\kappa_\mathcal{L}$ on the fine lattices.\hspace*{\fill}  }
		\label{fig:IRMatchingDifferentPhasesKappa}
	\end{subfigure}
	\caption{Results of the IR-Matching method for different fine lattice sizes $\mathcal{L}$ starting from a coarse lattice with $L=4$. The couplings on the fine lattice were fixed and only $\kappa_L(\theta)$ on the coarse lattice was optimisable. The bottom colorbars indicate the transition between phases by showing the $\log \xi_L/L$ for the x-axis' couplings.\hspace*{\fill} }
	\label{fig:IRMatchingDifferentPhases}
\end{figure*}
%

%%%%%%%%%%%%%%%%%%%%%%%%%%%%%%%%%%%%%%%%%%%%
\subsection{Results}
\label{sec:IRMatchingResults}

To put the IR-Matching method to work, we test its performance across the phase
diagram of a two-di\-mensional $\phi^4$-theory.
 
For each normalising flow, the coarsest grid is given by a two-dimensional
lattice with a linear lattice size of $L=4$. To gain a better intuition for the
evolution of the couplings, we only leave one coupling open for optimisation.
Specifically, we fix the on-site coupling $\lambda=0.01$ and only optimise the
hopping-term $\kappa_L(\theta)$. Nevertheless, optimising multiple couplings is
entirely feasible and unproblematic.
 
We also want to give an intuition for the phase transition and the related
correlations. Therefore, we show the logarithm of the correlation length
$\xi_L$ relative to the linear lattice size $L$ of the coarse lattice as a
function of $\kappa_\mathcal{L}$ in the colorbar at the bottom
\Cref{fig:IRMatchingDifferentPhasesESS} and
\Cref{fig:IRMatchingDifferentPhasesKappa}. The system is in the symmetric phase
for small values of $\kappa$ and enters the broken phase when increasing its
value.

In \Cref{fig:IRMatchingDifferentPhasesESS}, we show the ESS$/N$ after training the models to saturation. The shaded areas indicate the error of the ESS$/N$. As the latter is especially noisy in the broken phase, the errors are more fluctuant here. To begin with, the IR-Matching method works very well in the symmetric phase, even for lattices as fine as $128 \times 128$, obtaining an ESS/$N$ of well over 80\%. 
 
Near the critical region, we expect to see remnants of critical slowing down, since it mirrors the physics of the system. And, indeed we notice a drop in performance. Nevertheless, especially for lattices up to $64 \times 64$ the ESS/$N$ remains high enough not to harm practical use, steadily achieving ESS/$N$ over 60\%.
 
Moreover, even in the broken phase where normalising flows often fail due to mode collapse, the ESS$/N$ rarely drops below 40\% and for most lattices remains around or over 80\%. The ability of the flow to also work well in the broken phase, characterised by a multi-modal weight, is easily understood, as our base distribution is already multi-modal.
 
For comparison, we also plot the ESS$/N$ for a standard CNF for the $32 \times
32$ lattice as described in \Cref{sec:ArchitectureNormalizingFlow} but now with
full translational symmetry and a kernel that spans the entire system. We see
that it achieves comparable performance in the symmetric phase but is not
performant when entering the broken phase. We train the CNF as described in
\Cref{sec:HyperparametersAndErrors}, but adopt the hyperparameters ($F=30,
D=10, F'=20, D'=20$) using the notation established in
\Cref{sec:ArchitectureNormalizingFlow} and use a learning rate of $5\times
10^{-3}$ for stability. Notably, we significantly reduce the number of learnable
parameters. This originates in the fact that we can heavily restrict the CNF kernel
for the IR-Matching. In this case, the standard CNF for the $32 \times 32$ lattice has
approximately twice as many learnable parameters as the IR-Matching method. Moreover, 
for the next larger $64 \times 64$ lattice, the standard CNF has approximately five times as many learnable parameters.

Since the actions on the coarse and fine lattice roughly correspond to
different discretisations of the same continuum theory, we can relate well 
the couplings $\kappa_L(\theta)$ and $\kappa_\mathcal{L}$ on both levels. This
is shown in \Cref{fig:IRMatchingDifferentPhasesKappa}.

 \begin{figure}[t]
 	\centering
 	\includegraphics[width=0.45\textwidth]{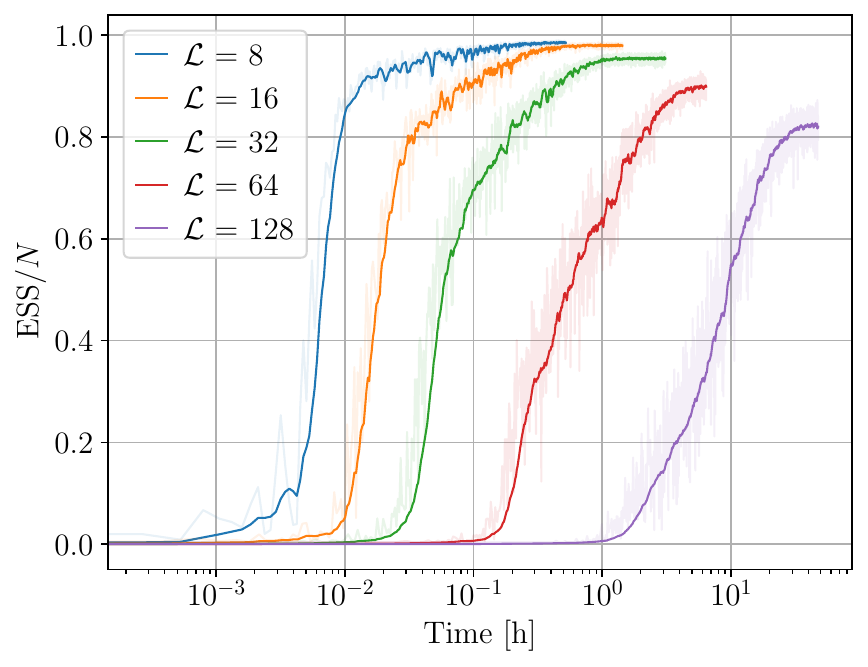}
 	\caption{ESS/$N$ for the IR-Matching method for different linear lattice sizes $\mathcal{L}$ of the target lattice. The bare coupling $\lambda_\mathcal{L}=0.01, \kappa_\mathcal{L}=0.25$ are used for all finer lattices, which brings the system into the vicinity of the critical region from the symmetric phase. The bold lines show the moving average whereas the shaded lines show the ESS$/N$ during training. \hspace*{\fill} }
 	\label{fig:IRMatching_ComparisonSystemSize_ESSTraining}
 \end{figure}
We display the coarse lattice couplings $\kappa_L(\theta)$ on the x-axis
together with $\log \xi_L/L$ to indicate the phase transition, while the y-axis
shows the fixed couplings $\kappa_\mathcal{L}$. The optimisable coarse
couplings cleanly sever the plot into two regions. These regions are related to the deeply
symmetric and broken phase. A gap in $\kappa_\mathcal{L}$ emerges close to the
(smeared out) phase transition. This means that for simulating physics close to
the transition on the finer lattice, the optimal coarse-level couplings lie in
the symmetric or broken phase but not close to the critical physics. This
behaviour is an advantage of the IR-Matching method, as simulating the symmetric
or broken phase on the coarse lattice is easier than simulating in the critical
region.
 
Relating the couplings is especially intuitive when we think of
$\kappa_L(\theta)$ as the lattice coupling after some coarse-graining steps
starting from the finer lattice at coupling $\kappa_\mathcal{L}$. The critical
point of the theory is an unstable fixed point with respect the to block
spinning transformation $\mathcal{T}_\theta^\dagger$ in coupling space.
Accordingly, applying successive RG steps drives the bare couplings away from
the critical point when starting in its vicinity on the fine level.
 
\vspace{\baselineskip}
To further stress the practical benefit of the IR-Matching method, we show the
training times for each lattice in
\Cref{fig:IRMatching_ComparisonSystemSize_ESSTraining}. The used hardware is
described in \Cref{sec:HyperparametersAndErrors}. We trained flows for
different lattice sizes up to $\mathcal{L}=128$, again starting from a coarse
lattice with $L=4$. For the bare couplings, we use $\kappa_\mathcal{L} = 0.25$
and $\lambda_\mathcal{L} = 0.01$, approaching the critical region from the
symmetric phase. The bold lines show the moving average of the ESS$/N$, whereas
the shaded lines indicate its momentary value.
 
We directly observe that the lattices with $\mathcal{L} \leq 32$ are trained
readily in under one hour and also systems like the $64 \times 64$ lattice are
trained in under seven hours. For the $128 \times 128$ lattices, we generally
see convergence of the ESS/$N$ after around 48 hours of training.
 
Since the IR-Matching introduces the coarse lattice as an optimisable object,
its training times and performance may be significantly improved by using finer
lattices to start with, offering a structural way to improve the flow, going
beyond making the learned map more expressive.

%%%%%%%%%%%%%%%%%%%%%%%%%%%%%%%%%%%%%%%%%%%%
%%%%%%%%%%%%%%%%%%%%%%%%%%%%%%%%%%%%%%%%%%%%
\section{Conclusion}
\label{sec:Conclusion}

In this paper, we have developed normalising flows for lattice field theories, 
inspired by the renormalisation group. We have proposed an
architecture, that is well suited to find transformations between a coarse and
a fine lattice with respect to a common action formula. Moreover, we also also 
introduced the \textit{IR-Matching} method, as a practical application, see \Cref{sec:IRMatching}. To our
knowledge this novel method outperforms current benchmarks for normalising flows regarding the
applicability in the broken phase and the extension to $128\times 128$ systems.
 
In the IR-Matching method from \Cref{sec:IRMatching}, we fix the couplings on
the finer lattice and optimise the couplings on the coarse lattice.
Furthermore, in \Cref{sec:UVMatching}, we also introduce the \textit{UV-Matching} method. There we we fix the couplings on the coarse lattice and aim to find an iterable
transformation to finer lattices, where the fine lattice couplings become
optimisable parameters.
 
In the application of the IR-Matching method we showed that one can efficiently sample a
$\phi^4$-theory in two dimensions on lattices as large as $128 \times 128$
while only requiring Monte Carlo configurations from a $4 \times 4$ system.
Moreover, we found that the couplings are driven away from the critical region
when coarsening. Conversely, this allows us to use samples from the symmetric
or broken phase on the coarse lattice to sample from the critical region on the
fine lattice. This is indeed expected from the renormalisation group and
further benefits the efficiency of the sampling process. Moreover, the training
times for the IR-Matching method are very reasonable, as for instance the $64
\times 64$ systems were trained in only a few hours.
 
Furthermore, just as for standard normalising flows, this approach instantiates
a rather general sampling technique for a lattice field theory, only requiring
plugging in the action and drift of the theory. Accordingly, the proposed
method is straightforwardly generalised to any other scalar models. Applications to long-ranged and fermionic theories~\cite{Attanasio:2024fte} are direct avenues for
future research. Moreover, since the sampling process naturally moves through different
resolutions, it potentially ties in nicely with multi-level Monte Carlo
approaches on the lattice~\cite{Jansen:2020}.
 
At last, we have discussed the capacity of our approach to combine the benefits
of traditional MCMC methods and normalising flows. However, since normalising
flows typically perform well on coarse lattices, one could first train a
standard coupling-conditional normalising flow on a very coarse discretization.
Then, this renormalisation group inspired flow may be used to upscale the
generated configuration, leading to a normalising flow all the way
down. This is left to future work.
 
By incorporating the coarse action into the  optimisation process, along with
the linear size of the coarsest lattice, we have introduced  a practical approach that
extends beyond making the machine learning model more expressive. This opens up
many new interesting and physics-informed avenues for normalising flows in 
lattice field theory and beyond. Specifically, we aim at a combination of the current framework with the physics-informed renormalisation group \cite{Ihssen:2024ihp} and the flow-based density of states \cite{Pawlowski:2022rdn}. We hope to report on these combinations in the near future.

%%%%%%%%%%%%%%%%%%%%%%%%%%%%%%%%%%%%%%%%%%%%

\section*{Acknowledgments}
We thank Friedericke Ihssen and Julian Urban for discussions and collaborations on related subjects. This work is funded by the Deutsche Forschungsgemeinschaft
(DFG, German Research Foundation) under
Germany’s Excellence Strategy EXC 2181/1 - 390900948
(the Heidelberg STRUCTURES Excellence Cluster) and
the Collaborative Research Centre SFB 1225 (ISOQUANT). We thank ECT* for
support at the Workshop `Machine Learning and the Renormalization Group' during
which this work has been developed further. The authors acknowledge support by the
state of Baden-Württemberg through bwHPC.
\appendix

%%%%%%%%%%%%%%%%%%%%%%%%%%%%%%%%%%%%%%%%%%%%
\section{UV-Matching}
\label{sec:UVMatching}

\begin{figure}[t]
	\centering
	\includegraphics[width=0.45\textwidth]{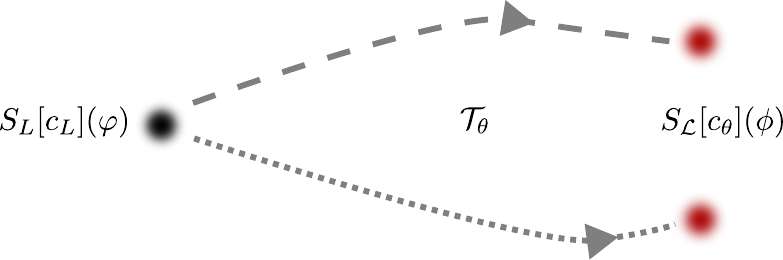}
	\caption{Illustration of the UV-Matching. The field $\varphi$ on the coarse lattice with action $S_L[c_L]$ and fixed couplings $c_L$ is connected to the field $\phi$ on the fine lattice with action $S_\mathcal{L}[c_\theta]$ and optimisable couplings $c_\theta$ by the transformation $\mathcal{T}_\theta$.\hspace*{\fill} }
	\label{fig:UVMatching}
\end{figure}
In \Cref{sec:IRMatching}, we have introduced the IR-Matching method. In this approach we kept the fine lattice couplings $c_\mathcal{L}$ constant and optimised
the coarse lattice couplings $c_L(\theta)$ as well as the transformation
$\mathcal{T}_\theta$. For the \textit{UV-Matching} as illustrated in
\Cref{fig:UVMatching}, we are interested in the opposite case: we keep
the coarse lattice couplings constant and optimise the transformation
$\mathcal{T}_\theta$ as well as the fine lattice couplings
$c_\mathcal{L}(\theta)$.

%%%%%%%%%%%%%%%%%%%%%%%%%
\subsection{The Method in Detail}
\label{sec:UVMatchingMethodExplanation}

\begin{figure*}[t]
	\centering
	\includegraphics[width=0.6\textwidth]{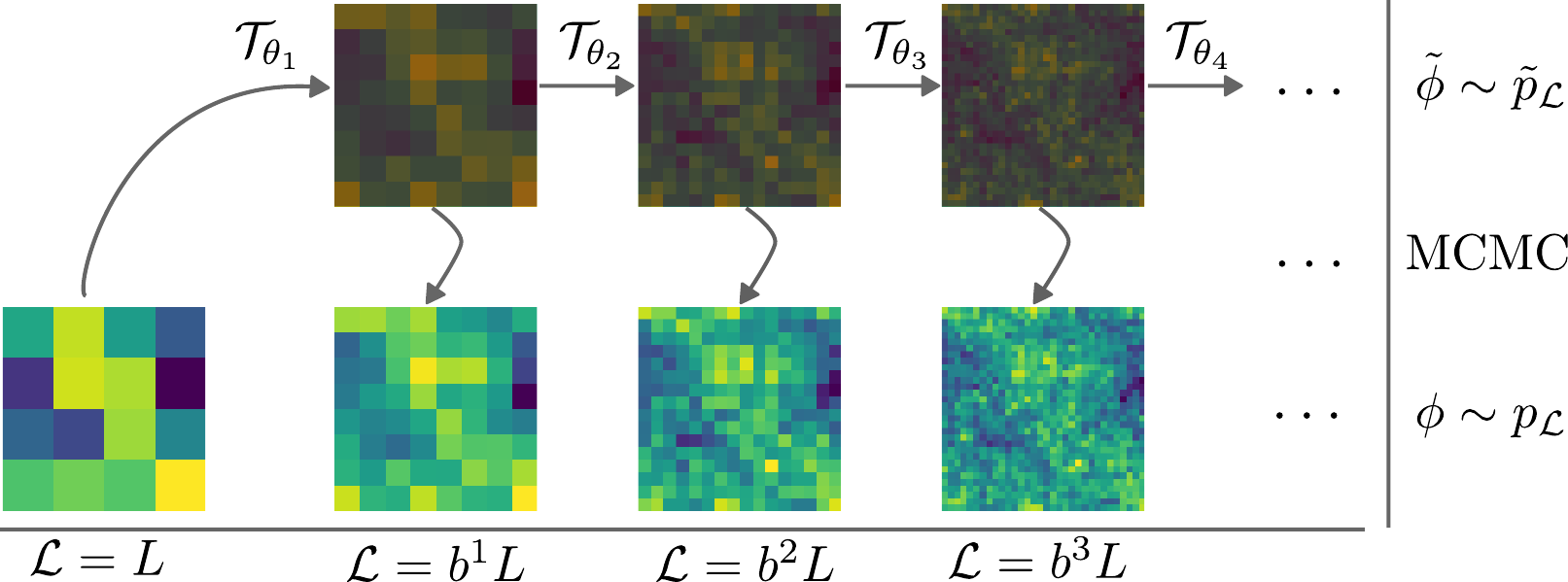}
	\caption{Iterative application of the learned flow $\mathcal{T}_{\theta_i}$ with learnable parameters $\theta_i$. After each application of the transformation one receives samples $\Tilde{\phi} \sim \Tilde{p}_\mathcal{L}$ with $\mathcal{L} = b^i L$. Then, a MCMC accept reject step from \Cref{eq:AcceptReject} can be conducted to ensure that one samples from the $\phi^4$-theory distribution $p_\mathcal{L}$ on the finer lattice.\hspace*{\fill} }
	\label{fig:UVMatchingIterativeApplication}
\end{figure*}
To begin with, the general training procedure is very similar to the one
described for the IR-Matching. We sample coarse configurations $\varphi$, push
them through the transformation $\mathcal{T}_\theta$. Then we minimise the
Kullback-Leibler divergence from \labelcref{eq:ReverseKLDivergence}, but in contradistinction to the IR-Matching method we 
leave the fine lattice couplings open for optimisation. Accordingly, we do
not just optimise the overlap between the target and push forward distribution
by changing the latter. Instead we allow for the target distribution to change
slightly in the direction of the current push forward.
 
As the coarse lattice couplings are now fixed, we can use a traditional HMC
algorithm to draw samples from the coarse distribution. However, optimising the
couplings on the fine lattice requires us to draw configurations of the
respective system during training to estimate the gradient from
\labelcref{eq:logZGradient}.
 
This has the potential to become computationally expensive and to re-introduce
critical slowing down. It is here, where the reweighing approach from
\labelcref{eq:Reweighing} turns out to be very useful. Since the configurations used to
compute the $\log Z$ gradient are not pushed through the network, we can safely
reuse them. By using reweighing the need to draw samples from the target
distribution at every learning step is reduced, significantly lowering the
number of Monte Carlo samples required for training. Note that reweighing is
only possible if the coupling changes slowly. One could in principle also use
configurations from the machine learning model during retraining steps to
estimate the $\log Z$ gradient. However, this is left to future work. 
 
Notably, the learned transformation $\mathcal{T}_{\theta}$ does not explicitly
depend on the linear lattice size. This is straight--forwardly true for the
na\"ive upsampling. Furthermore, the system size independence
also applies for the rest of the architecture in \Cref{fig:FlowExplained}, as the addition of noise is conducted per
upsampling block $\mathcal{B}$ and the normalising flow learns a kernel
$\mathcal{K}$ of fixed size across the lattice. 
 
Accordingly, one can learn the transformation $\mathcal{T}_\theta$ on a coarse
lattice where training and sampling is efficient. Then one applies this
transformation iteratively many times to finer lattices. This iterative
application is illustrated in \Cref{fig:UVMatchingIterativeApplication}, where
we start from a coarse lattice with linear lattice size $L$ and double the
linear lattice size after each application, $\mathcal{L}= b^iL$.
 
In spirit this is very similar and indeed inspired by the works
in~\cite{Bachtis:2021eww,SR-Ising}, where the authors use convolutional neural
networks. The great benefit of our approach is that we can efficiently keep
track of the log--likelihoods of the samples. Firstly, this enables the direct
application of Monte Carlo methods to make the approach exact, see
\Cref{fig:UVMatchingIterativeApplication}. Secondly, it also allows us to
directly monitor the efficiency of the method, which is indicated by the acceptance rate. 
This allows one to make an educated assertion of the maximal number of
applications of the transformation.

%%%%%%%%%%%%%%%%%%%%%%%%%%%%%%%%%%%%%%%%%%%%
\subsection{Results}
\label{sec:UVMatchingMethodCouplings}

The training only directly determines the optimal coupling after the first
iteration of the transformation $\mathcal{T}_{\theta_1}$, where we double the
linear lattice size of the coarsest field. As we want to apply the
transformation iteratively, we require a sensible approach to choose the
optimal coupling after further iterations. Here, we straight--forwardly choose
the couplings that maximise the ESS$/N$ on the respective lattice.
 
Showing the estimation of the coupling explicitly is especially illustrative in
the one-dimensional case: each iteration only doubles but not quadruples
the total number of lattice sites and most changes vary less strongly than in
higher dimensions. We thus match a one-dimensional $\phi^4$-theory on a lattice
with linear lattice size $L=8$ and couplings $\kappa_L=0.1, \lambda_L=0.01$ to
a lattice with linear lattice size $\mathcal{L}=16$, fixed coupling
$\lambda_\mathcal{L}=\lambda_L$, and optimisable coupling
$\kappa_\mathcal{L}(\theta)$.
 
\begin{figure}[t]
	\centering
	\includegraphics[width=0.45\textwidth]{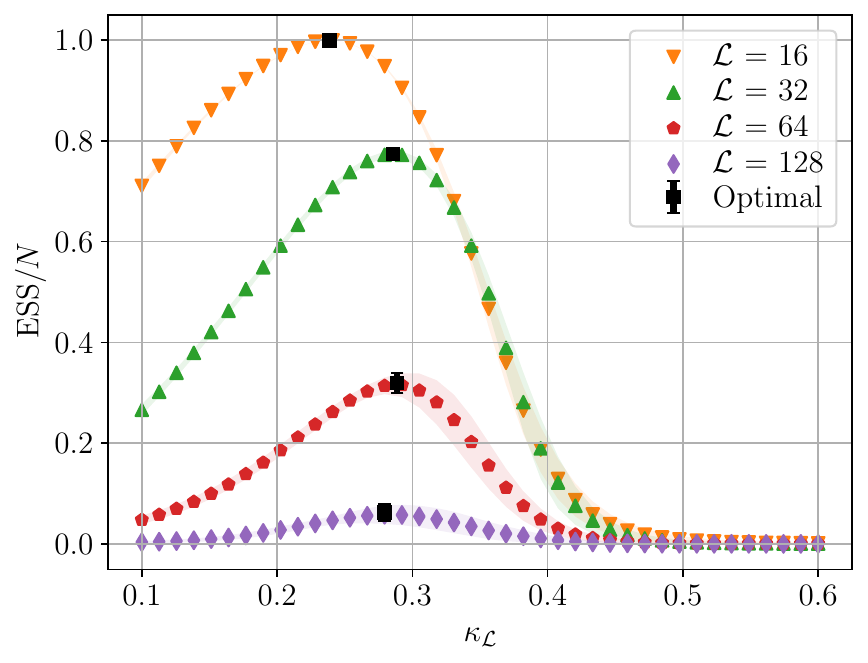}
	\caption{Estimation of the coupling $\kappa_\mathcal{L}(\theta)$ for a one-dimensional $\phi^4$-theory by optimising the ESS/$N$ after iterative application of the UV-Matching flow.\hspace*{\fill} }
	\label{fig:UVMatchingCouplingEstimation}
\end{figure}

In \Cref{fig:UVMatchingCouplingEstimation}, we show the ESS/$N$ for different
couplings after iterative application of the flow. We firstly note that the
ESS/$N$ varies smoothly, making it easy to estimate the optimal value. While we
see that the straight--forward iteration of the learned flow works, it is not
apt to reach finer lattices efficiently as the ESS$/N$ drops approximately
exponentially.

To improve performance, we note that the straight-forwardly iterated
transformation may not be optimal, but should also not be too far away from the
optimal transformation. Accordingly, only a few retraining steps $N_{RT}$ can
suffice to obtain good results. For retraining, we again sample with the flow
on the respective fine lattice we are currently optimising and minimise the KL
divergence from \labelcref{eq:ReverseKLDivergence} with a learning rate that we
reduced by a factor of five. Other than that everything else stays the same.
 
\begin{figure}[t]
	\centering
	\includegraphics[width=0.45\textwidth]{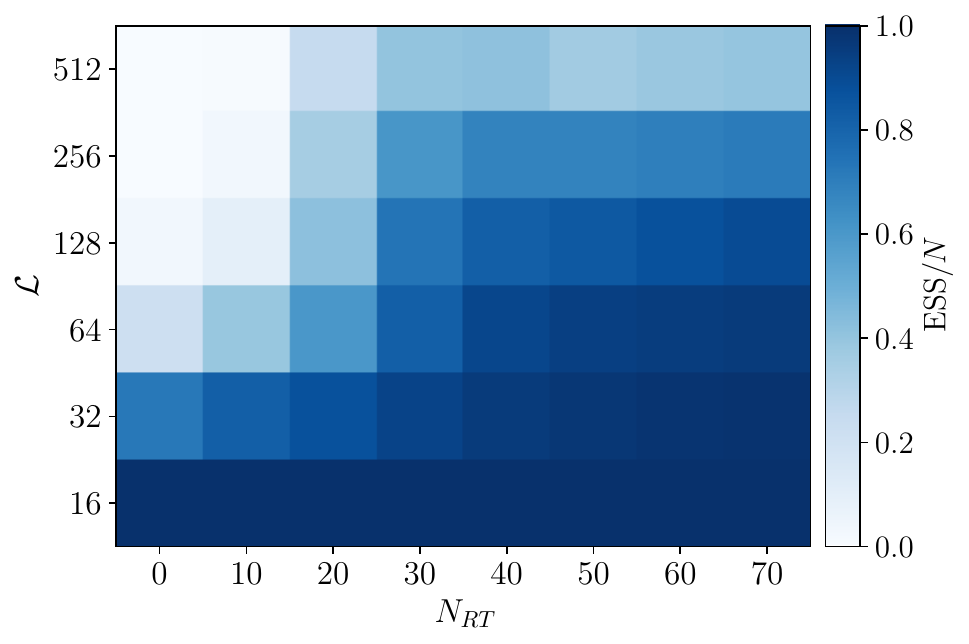}
	\caption{The optimal ESS$/N$ for the learned flow from \Cref{fig:UVMatchingCouplingEstimation} after $N_{RT}$ retraining steps.\hspace*{\fill} }
	\label{fig:UVMatchingRelearning}
\end{figure}

In \Cref{fig:UVMatchingRelearning}, we provide the optimal ESS$/N$ for the
learned flow after applying $N_{RT}$ retraining steps. As can be seen, we reach
good performance up to lattices with $\mathcal{L}=512$ for the one-dimensional
$\phi^4$-theory and the ESS$/N$ saturates after already $50$ retraining steps.
In practice, we use early--stopping, checking whether the ESS has saturated or
not, to determine the optimal number of retraining steps.

We now examine the behaviour of the UV--Matching method across different phases
on lattices of varying sizes, and discuss how it is related to the 
physics of the system. Here, we again return to the two-dimensional lattice. Specifically, we
fix coarse lattices in both the symmetric and broken phases near the critical
region, and aim to optimise the transformations $\mathcal{T}_{\theta_i}$.

\begin{figure*}[t]
	\centering
	\begin{subfigure}[t]{0.45\textwidth}
		\centering
		\includegraphics[width=\textwidth]{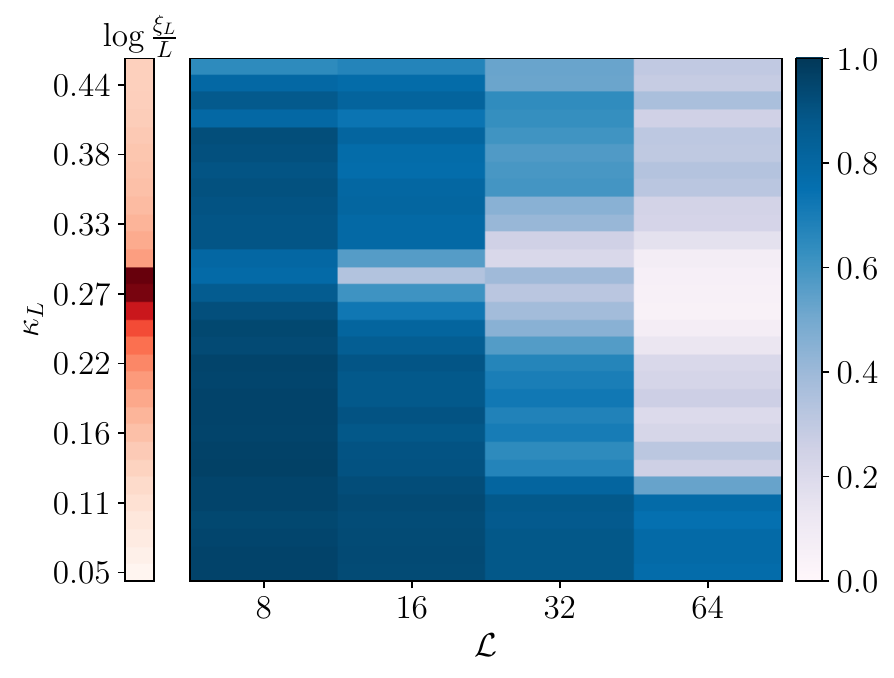}
		\caption{ESS$/N$ for the UV--Matching for different linear lattice sizes $\mathcal{L}$ across the phase structure of the system. \hspace*{\fill} }
		\label{fig:UVMatchingDifferentPhasesESS}
	\end{subfigure}
	\hspace{0.05\textwidth}
	\begin{subfigure}[t]{0.45\textwidth}
		\centering
		\includegraphics[width=\textwidth]{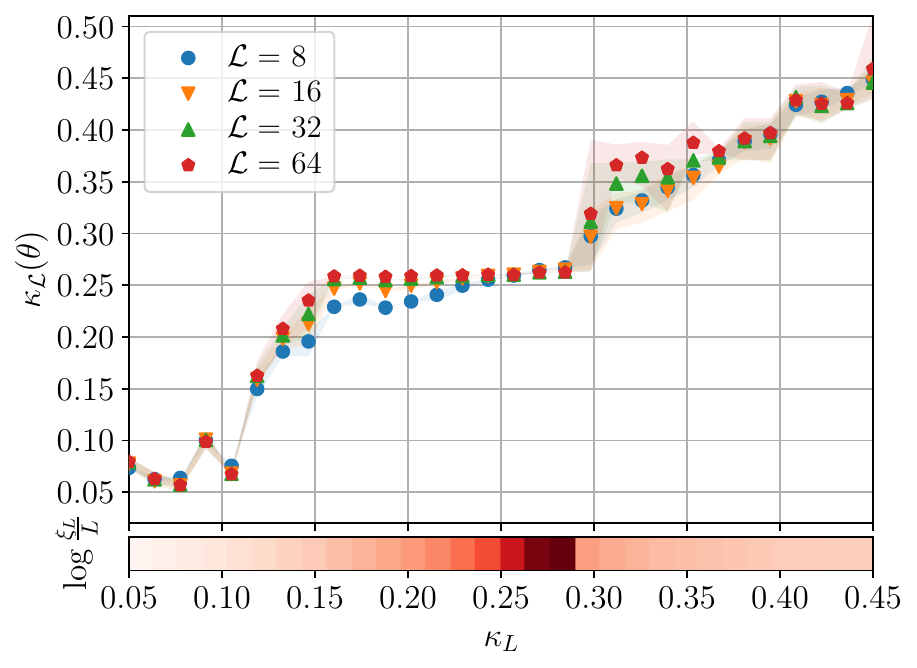}
		\caption{Optimizable coupling $\kappa_\mathcal{L}(\theta)$ for the UV--Matching.\hspace*{\fill} }
		\label{fig:UVMatchingDifferentPhasesKappa}
	\end{subfigure}
	\caption{Results for the UV-Matching method for different number of iterations of the transformation $\mathcal{T}_\theta$ starting from a coarse lattice with $L=4$. The couplings on the coarse lattice were fixed and only $\kappa_\mathcal{L}(\theta)$ on the fine lattice was optimisable. To indicate the transition between phases, we display colorbars for $\log \xi_L/L$ on the coarse lattice.\hspace*{\fill} }
	\label{fig:UVMatchingDifferentPhases}
\end{figure*}

In \Cref{fig:UVMatchingDifferentPhasesESS}, we plot the ESS$/N$ after each iteration of the flow to lattices with a linear size $\mathcal{L}$. The starting couplings are taken from the coarse lattice, as shown on the y-axis, along with the respective values of $\log \xi_L/L$ to provide context for the critical region. First, we observe that the ESS$/N$ remains high in almost every case for lattices with sizes up to $\mathcal{L} = 32$. Even on the $64 \times 64$ lattice, the ESS$/N$ remains above 50\% in both the deeply symmetric and broken phases. However, starting from the critical region on the coarsest lattice, we notice a cone of decreasing ESS$/N$ when moving to finer grids.

To make sense out of this, we show the learned couplings relative to the couplings
on the coarsest lattice in \Cref{fig:UVMatchingDifferentPhasesKappa}. In
contrast to the IR-Matching, the couplings actually move towards their critical
values when applying the inverse block spinning transformation iteratively in
the vicinity of a phase transition. This originates in the fact that we now optimise the couplings on the
finer lattice. Accordingly, as sampling and training on the critical point
posits a computationally challenging problem, the ESS$/N$ drops significantly
in each iteration.
 
When comparing the behaviour of the couplings for the IR- and UV-Matching in
\Cref{fig:IRMatchingDifferentPhasesKappa} and
\Cref{fig:UVMatchingDifferentPhasesKappa} respectively, we see the 
plateaus for the critical coupling $\kappa_\mathcal{L}$. For the IR-Matching method, this
plateau manifests as a gap, where the coarse lattice couplings
move away from the plateau. For the UV-Matching method, the couplings actually move
towards it.

Accordingly, while the UV-Matching method should not be understood as a method
to simulate the whole parameter set of the field theory, it lends itself to
simulating critical physics on the lattice.

%%%%%%%%%%%%%%%%%%%%%%%%%%%%%%%%%%%%%%%%%%%%
\section{Relation to Block Spinning transformations}
\label{sec:BlockSpinningTransformation}

In the main text, we introduced renormalisation group inspired normalising
flows. The connection between the actions on the coarse and fine lattice ($S_L,
S_\mathcal{L}$) can be made explicit when we assume that the action on the fine
lattice is given.
 
In the following, we will call a transformation $\mathcal{T}_\theta^\dagger:
\mathds{R}^{\mathcal{L}^d} \to \mathds{R}^{L^d}$ a \textit{block spinning
transformation} when it does not violate any symmetries of the original lattice
and for the induced blocking.
\begin{align}
	P(\varphi|\phi) := \delta\left[ \varphi - \mathcal{T}^\dagger(\phi) \right]
\end{align}
it is true that
\begin{align}\nonumber 
	P(\varphi|\phi) \geq &\, 0 \quad \forall \phi,\varphi\,,\\[1ex]
	\int \mathcal{D}\varphi\, P(\varphi|\phi) =&\, 1 \quad \forall \phi\,.
	\label{eq:ConditionsBlockSpinning}
\end{align}
Then, when $\phi$ follows the above-mentioned Boltzmann distribution, the block
spinning transformation also induces a Boltzmann distribution for the coarser
field $\varphi$ via
\begin{align}
	e^{-S_L(\varphi)} = \int D\phi\, P(\varphi| \phi)\, e^{-S_\mathcal{L}(\phi)}\,.
\label{eq:InducedBlockSpinningMeasure}
\end{align}
This construction then ensures that the transformation leaves the partition sum
invariant, as is directly seen by
\begin{align}\nonumber 
	Z_L = &\,\int D\varphi\, e^{-S_L(\varphi)} \\[1ex]\nonumber 
	 \overset{\eqref{eq:InducedBlockSpinningMeasure}}{=}&\, \int D\varphi\, D\phi\, P(\varphi, \phi)\, e^{-S_\mathcal{L}(\phi)}\\[1ex]
	\overset{\eqref{eq:ConditionsBlockSpinning}}{=}&\,\int D\phi\, e^{-S_\mathcal{L}(\phi)} = Z_\mathcal{L}\,.
\end{align}
Given the context of \labelcref{eq:InducedBlockSpinningMeasure}, we now want to
consider how the actions on the coarse and fine lattice are connected to each
other when we consider $\mathcal{T}_\theta^\dagger$ as a block spinning
transformation. Starting from the fine action and applying
\labelcref{eq:InducedBlockSpinningMeasure} directly leads us to
\begin{align}
    \label{eq:BlockSpinningRelation_A}
	e^{-S_L(\varphi)} = \int D\phi\, \delta\left[ \varphi - \mathcal{T}^\dagger_\theta(\phi) \right]\, e^{-S_\mathcal{L}(\phi)} 
\end{align}
Now, we would like to connect the fine field $\phi \in
\mathds{R}^{\mathcal{L}^d}$ to the coarse field $\varphi \in \mathds{R}^{L^d}$
in a more explicit manner. To this end, we revert the steps indicated in
\Cref{fig:FlowExplained} and, hence, apply the normalising flow $F_\theta$
inversely. Under the integral, we now call $\Psi = F_\theta^{-1}(\phi)$ and
using \labelcref{eq:FlowApplication} together with
\labelcref{eq:BlockSpinningRelation_A}, we obtain
\begin{align}\nonumber 
	    e^{-S_L(\varphi)} = &\,\int D\Psi \, \delta\left[ \varphi - \mathcal{T}^\dagger_\theta\circ F_\theta (\Psi) \right]
	     \\[1ex]
	&\hspace{1.2cm}\times\, e^{-S_\mathcal{L}(\phi) +\log \det J_F(\Psi)}\,.
 \label{eq:BlockSpinningRelation_B}
\end{align}
Next, we note that we can express each variable $\Psi$ according to
\labelcref{eq:InvertibleNoiseAddition} and \labelcref{eq:NaiveUpsampling} as
$\Psi = U(\varphi') + \zeta$. More precisely, we now only consider the noise
degrees of freedom $\zeta' \in \mathds{R}^{\mathcal{L}^d-L^d}$ and can connect
them to the full noise field via \labelcref{eq:NoiseConstraint}. We do so
because now we may define the field $\psi' = (\varphi', \zeta')^T$ that has the
same dimension as $\Psi$, and we can rewrite it as
\begin{align}
	\Psi = U_\psi(\psi') = U(\varphi') + V(\zeta')\,.
\label{eq:BlockSpinningRelation_psiPsi}
\end{align}
We can equally rewrite the integration over $\Psi$ as an integration over
$\psi'$ or $\varphi'$ and $\zeta'$ respectively. The respective Jacobian
$J_{U_\psi}$ of this transformation is field-independent as $U$ and $V$ are
linear transformations relating to \labelcref{eq:NaiveUpsampling} and
\labelcref{eq:NoiseConstraint} respectively. So, this transformation only picks
up a constant factor. Accordingly, we obtain from
\labelcref{eq:BlockSpinningRelation_B,eq:BlockSpinningRelation_psiPsi}
\begin{align}\nonumber 
	e^{-S_L(\varphi)} \simeq &\,\int D\varphi'\, D\zeta'\, \delta\left[ \varphi - \mathcal{T}^\dagger_\theta \circ F_\theta \circ U_\psi(\, \psi' \,) \right]\\[1ex]
	&\hspace{1.8cm}\times e^{-S_\mathcal{L}(\phi) +\log \det J_F(\Psi)} \,.
\end{align}
Now, we can see that $F_\theta \circ U_\psi$ is precisely the transformation
$\mathcal{T}_\theta$ for a fixed noise field described in
\Cref{fig:FlowExplained}, leading to the further simplification
\begin{align}\nonumber 
	e^{-S_L(\varphi)} \simeq &\,\int D\varphi'\, D\zeta'\, \delta\left[ \varphi - \mathcal{T}^\dagger_\theta \circ \mathcal{T}_\theta(\varphi'; \zeta') \right] \\[1ex]
	&\hspace{1.8cm} \times  e^{-S_\mathcal{L}(\phi) +\log \det J_F(\Psi)} \,.
  \label{eq:BlockSpinningRelation_C}
\end{align}
Accordingly, we can use the left inverse property of the transformation
$\mathcal{T}_\theta$ from \labelcref{eq:LeftInverse}, such that
$\mathcal{T}^\dagger_\theta \circ \mathcal{T}_\theta(\varphi'; \zeta') =
\varphi'$. This enables us to perform the integration over $\varphi'$, and we
obtain from \labelcref{eq:BlockSpinningRelation_C}
\begin{align}
    \label{eq:BlockSpinningRelation_Result}
	\begin{aligned}
		e^{-S_L(\varphi)} \simeq \int D\zeta'\, e^{-S_\mathcal{L}(\phi) +\log \det J_F(\Psi)}\,.
	\end{aligned}
\end{align}
As in the equations before, we have kept the implicit notation, such that $\phi
= \mathcal{T}_\theta(\varphi; \zeta')$ and $\Psi = U(\varphi) + V(\zeta')$
under the integral. The result from \labelcref{eq:BlockSpinningRelation_Result}
shows nicely, that the block spinning transformation as implemented by
$\mathcal{T}^\dagger_\theta$ literally boils down to an integration over the
noise degrees of freedom. Moreover, it lets us explicitly connect the coarse
action $S_L$ to the fine action $S_\mathcal{L}$.

%%%%%%%%%%%%%%%%%%%%%%%%%%%%%%%%%%%%%%%%%%%%
\section{Hyperparameters and error analysis}
\label{sec:HyperparametersAndErrors}

In this Appendix we describe the hyperparameters used for our architecture.
 
For the na\"ive upsampling, there are no hyperparameters present. Furthermore,
for the addition of noise, at the beginning of the training, we initialise the
learnable variance $\sigma_\theta^2$ of the Gaussian noise $\zeta$ to equal the
variance of the configurations $\varphi$ on the coarse lattice. This is done to
ensure that the noise does not dominate the flow at the beginning of the
training.
 
To compute the $\log Z$ gradients, we sample $3k$ configurations on the
respective coarse lattice with a standard Hybrid Monte Carlo (HMC) algorithm.
For thermalisation, we start from an i.i.d initial Gaussian configuration and
run the Hybrid Monte Carlo Code until $2k$ proposals have been accepted. We
save a sample after each HMC step.
 
Since reweighing is used in this part of training, we have to choose a point at
which we want to sample new configurations. For this, we also store the
likelihood of the configurations when they are sampled. The ESS/$N$ for each new
coupling is then computed, and we choose a lower bound $R_W$ for the ESS/$N$
upon which we sample anew. In our experience, $R_W = 0.85$ gives a good
compromise between speed and accuracy.
 
For the IR-Matching from \Cref{sec:IRMatching}, we use a batched Langevin
Sampling algorithm from \labelcref{eq:Langevin} on the coarsest lattice. We use
a step size of $\tau = 0.01$ for this. Training starts from an i.i.d. initial
Gaussian configurations and involves thermalising the system for $5\times 10^4$
steps. Between each training step, we re-thermalise the system by taking $500$
Langevin steps, starting from the current batch of configurations.
 
For the CNF, we achieved good results when using $F=11$ learnable sine
frequencies and choosing $D=10$ for the kernel number. The time and frequency
space bond dimensions are $F'=20$ and $D'=20$, respectively. In all
computations, the coarsest lattice of the flow has a linear lattice size of
$L=4$ to showcase that one only needs very coarse lattices. Hence, the kernel
size is also limited. We found a kernel size of $2$ sufficient for the
IR-Matching as presented here and used a kernel size of $3$ for the UV-Matching.
 
For optimisation, we use the Adam optimiser with decaying parameters
$\beta_1=0.8$ and $ \beta_2 =0.9$ as suggested in \cite{Gerdes:2023}. The initial
learning rate is set to $0.01$ and decays exponentially after each learning
step by a factor of $0.997$.
 
To estimate the errors for the ESS$/N$ and the optimised couplings, we trained
each model three times with different random seeds. The data point is then
given by the respective mean value, and the error bars reach the maximally and
minimally obtained value.
 
For the training and evaluation we used the resources of the bwHPC cluster.
Virtually all the trainings were conducted on a system with an NVIDIA Tesla
V100 GPU with 32 GB of memory, supported by Intel Xeon Gold 6230 processors (2
sockets, 40 cores total), and 384 GB of main memory. For the larger 128$\times$
128 lattice, we switched to a system with an NVIDIA A100 GPU with 80 GB of
memory, supported by Intel Xeon Platinum 8358 processors (2 sockets, 64 cores
total), and 512 GB of main memory. The latter switch was done to conveniently
accommodate the memory requirements of the larger lattice sizes.

%%%%%%%%%%%%%%%%%%%%%
\small
\bibliography{refs}

\end{document}